\documentclass[pre,aps,nofootinbib,twocolumn,showpacs]{revtex4}
\usepackage{graphics}
\usepackage{epsfig}

\begin{document}

\title{The phase transition of triplet reaction-diffusion models}
\author{G\'eza \'Odor}
\affiliation{Research Institute for Technical Physics and Materials Science, \\
H-1525 Budapest, P.O.Box 49, Hungary}

\begin{abstract}
The phase transitions classes of reaction-diffusion systems with
multi-particle reactions is an open challenging problem. Large scale 
simulations are applied for the $3A\to 4A$, $3A\to 2A$ and the
$3A\to 4A$, $3A\to\emptyset$ triplet reaction models with site 
occupation restriction in one dimension. 
Static and dynamic mean-field scaling is observed with signs of 
logarithmic corrections suggesting $d_c=1$ upper critical dimension
for this family of models.
\end{abstract}
\maketitle

\section{Introduction}

The exploration of the universal behavior of non-equilibrium phase
transitions is still an open problem of statistical physics 
\cite{Dick-Mar,Hin2000,dok}. In equilibrium critical phenomena 
symmetries, dimensions and boundary condition are the relevant factors 
determining a universality class. During the study of non-equilibrium
models other circumstances, like initial conditions or topological 
effects in low dimensions have been proven to be decisive \cite{OdMe02}.
Unfortunately solid field theoretical background exists for only a few
reaction-diffusion (RD) systems exhibiting continuous phase transition
to absorbing states \cite{Uweof}. These are mainly branching and 
annihilating random walk models $A\to (n+1)A$, $2A\to\emptyset$ 
built up from unary particle creation reactions 
\cite{Cardy-Tauber,HCDMlett,CCDDM05} RD models are interesting since many
other types of systems like surface growth, spin systems or 
stochastic cellular automata can be mapped on them \cite{dok}.

Recently RD systems with multi-particle creation has became in
the focus of research. Numerical studies resulted in debated critical
phenomena and generated a long series of publications. 
An interesting example was being investigated during the past decade
that emerges at phase transitions of binary production systems \cite{PCPD}
(PCPD). In these systems particle production competes with pair
annihilation and single particle diffusion. If the production wins
steady states with finite particle density appear in (site restricted) 
models with hard-core repulsion, while in unrestricted (bosonic)
models the density diverges. If the annihilation is stronger an 
absorbing phase emerges, which is either completely empty or 
contains a solitary diffusing particle. In between the two phases 
a continuous phase transition can be observed in site restricted models.

In triplet reaction models at least three particle is needed to contact
for a reaction. They have been investigated by simulations 
\cite{PHK02,KC0208497,tripcikk} and numerical Langevin equation
solution \cite{DCM05}. The first simulation results
\cite{PHK02,KC0208497} for $3A\to 4A$, $3A\to 2A$ models in one
dimension claimed a distinct universal behavior from the known ones.
Simple power-counting analysis of an effective Langevin equation corresponding
to the coarse grained microscopic model results in $d_c=4/3$
\cite{tripcikk}. However the numerical estimates for the critical
exponent describing the density decay from homogeneous random initial
state $\rho(t) \propto t^{-\alpha}$ differed significantly
$\alpha=0.32(1)$ \cite{PHK02} vs $\alpha=0.27(1)$ \cite{KC0208497}.
In the former case a site restricted model was investigated and
scaling was reported in the $10^4 < t < 10^6$ region. In \cite{KC0208497}
different suppressed bosonic triplet models -- where the multiple site 
occupancy is suppressed by an exponential probability factor --
exhibited scaling for $10^4 < t < 10^7$ Monte Carlo steps (MCS) 
(throughout the paper the time is measured by MCS).
Renormalization group analysis pointed out \cite{JWDT04} that a single
field theory does not exhibit a nontrivial stable fixed point and suggested
$d_c=1$ for such models. This study raised the possibility that a
proper field theory should be a coupled one, with positively correlated
clustered particles and solitary random walkers.
An other simulation study \cite{tripcikk} on site restricted models
reported scaling agreeing with mean-field exponents $\alpha=0.95(5)$
and $\beta=1.07(10)$, where $\beta$ is the order parameter exponent in
the active phase $\rho \propto |p-p_c|^{\beta}$.

Very recently a coupled field theoretical description of such systems
is suggested \cite{DCM05}. An effective Langevin equation between a DP
like and an annihilating random walk (ARW) system is analyzed by numerical 
integration technique. Note however the ARW like system is described
by a positive noise term saying that at the critical point the
anti-correlations do not play a role and serve merely as a fluctuating
source to the primary field. As the consequence this field theory
leads to the same critical scaling behavior as that of the PCPD albeit
with a different upper critical dimension $d_c=4/3$ (vs $d_c=2$ for PCPD).
So according to this study in one dimension one should see PCPD class exponents:
$\alpha=0.20(1)$ and $\beta = 0.40(1)$ \cite{KC0208497,pcpd2cikk}.

In the present study I extend the simulation time of the 
$3A\to 4A$, $3A\to 2A$ model investigated in \cite{tripcikk}
by two orders of magnitude and follow the static and dynamic scaling
behavior of particles and triplets ($AAA$) at different diffusion probabilities.
In Section (\ref{3430sec}) I apply this the same kind of analysis for the
$3A\to 4A$, $3A\to\emptyset$ model.

\section{Simulations of the $3A\to 4A$, $3A\to 2A$ model} \label{1dsimu}

The simulations were carried out on $2\times L=10^5$ sized systems
with periodic boundary conditions. The initial states were randomly
half filled lattices, and the density of particles, singlets and triplets
is followed up to $2\times 10^9$ MCS by random sequential dynamics.
An elementary MCS consists of the following steps. A particle $A$
is chosen randomly and the following processes are done:
\begin{description} 
\item[(a)] $A\emptyset\leftrightarrow\emptyset A$ with probability D,
\item[(b)] $3 A \to 2 A$ with probability $p(1-D)$,
\item[(c)] $3 A \to 4 A$ with probability $(1-p)(1-D)$,
\end{description}
such that the reactions were allowed on the left or right side of the 
selected particle strings randomly. The time is updated by $\rho(t)$.
The time -- measured by MCS --- is updated by $1/n_P$, where $n_P$ is the total particle number at time $t$. 
In order to get precise $\rho(t)$, critical point and exponent estimates
the number of independent realizations varied between 20 and 350
per $p$ and $D$ throughout this study. 

\begin{figure}
\epsfxsize=70mm
\epsffile{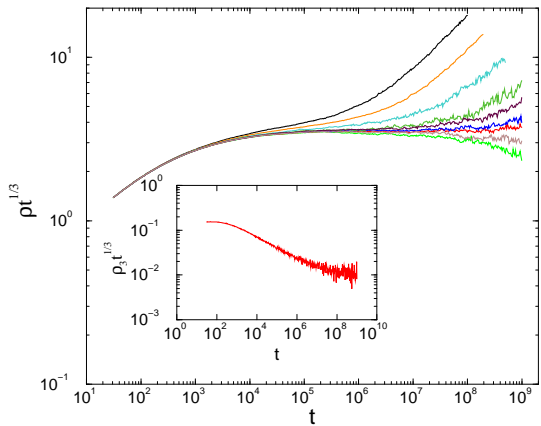}
\vspace{2mm}
\caption{(Color online)
Density decay of the $3A\to 4A$, $3A\to 2A$ model at $D=0.1$.
Different curves correspond to $p=0.301$, $0.302$, $0.3027, $$0.3031$, 
$0.3032$, $0.3033$, $0.30335$, $0.3034$, and $0.3035$ 
(from top to bottom). The insert shows the decay of triplets:
$\rho_3t^{1/3}$, for $p=0.30337$}
\label{3432dec_1}
\end{figure}

First I extended the simulations at $D=0.1$ published in
\cite{tripcikk} from $t_{max}=10^7$ MCS by a factor of 200
in time. Figure \ref{3432dec_1} shows the density decay multiplied by 
$t^{1/3}$.
Following an long initial transient, where the decay is slow an excellent
agreement with the mean-field scaling can be observed for 
$10^6 < t < 10^9$ MCS for $p_c= 0.30337(2)$. The triplet density
decays in the same way ($\rho_3\propto t^{1/3}$) as the total density
suggesting $d_c=1$.

I repeated the simulations for a higher diffusion rate $D=0.8$. In this
case due to the dynamics of this model the reaction rates are smaller
and one can observe a faster than mean-field decay for 
$10^4< t < 5\times 10^8$ MCS. Agreement with mean-field scaling sets 
in for $5\times 10^8 < t <2\times 10^9$ MCS steps for $p=0.4075(1)$.
Fast transient decay for intermediate times has already been explained
in case of the $AA\to A$ coagulation model with finite reaction rates
\cite{BMGR,PDF}. One may expect similar behavior for the triplet
annihilation case where the low reaction rates at $D=0.8$ can explain 
the fast transient seen here.
Again the triplet density decays by the mean-field law for 
$t>5\times 10^8$ MCS as expected at the upper critical dimension
$d_c=1$.
\begin{figure}
\epsfxsize=70mm
\epsffile{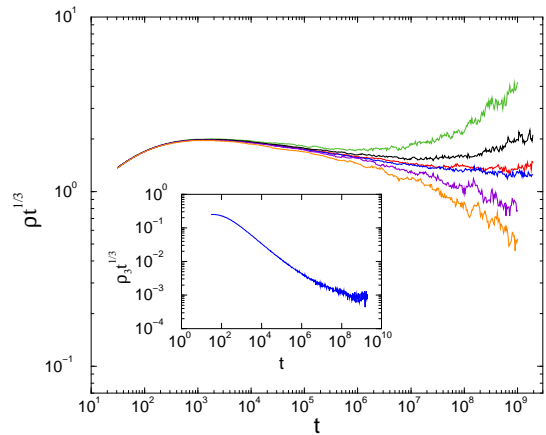}
\vspace{2mm}
\caption{(Color online)
Density decay of the $3A\to 4A$, $3A\to 2A$ model at $D=0.8$.
Different curves correspond to $p=0.406$, $0.407$, $0.4074$, $0.4075$ 
$0.408$, $0.409$ (from top to bottom). The insert shows the triplet
density decay at $p=0.4075$.}
\label{3432ins_8}
\end{figure}
Using the critical point estimates obtained by the dynamical simulations
I investigated the singular behavior of the order parameter in the
supercritical region as well. The order parameter is expected to scale
as
\begin{equation}
\rho \propto |p-p_c|^{\beta}
\end{equation}
To get the steady state densities I followed the decay in several samples
until saturation occurs and averaged it in a long time window
exceeding a level-off seen on log-lin. scale. The data are analyzed by
the local slopes method. According to this
\begin{equation}
\beta_{eff}(p_i) = \frac {\ln \rho(\infty,p_i) -
\ln \rho(\infty,p_{i-1})} {\ln(p_i) - \ln(p_{i-1})} \ \ ,
\label{beff}
\end{equation}
and as $p_i\to p_c$ the effective exponent tends to the true critical value
$\beta_{eff} \to \beta$.
\begin{figure}
\epsfxsize=70mm
\epsffile{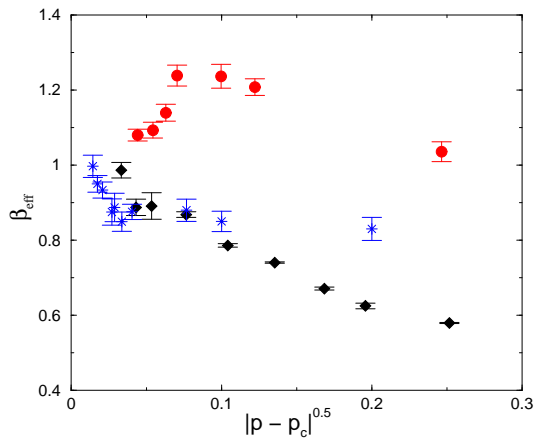}
\vspace{4mm}
\caption{(Color online)
Effective static order parameter exponent results in the active phase.
Bullets correspond to $D=0.8$, boxes to $D=0.1$ of the 
$3A\to 4A$, $3A\to 2A$ model. Stars denote the results of the 
$3A\to 4A$, $3A\to\emptyset$ model.}
\label{beta3432_18}
\end{figure}
As Figure \ref{beta3432_18} shows the $\beta_{eff}$ converges to the
mean-field value $\beta=1$ for both diffusion rates I investigated. 
For $D=0.8$ on can see an overshooting, while for $D=0.1$ local slopes 
approach the asymptotic value from below following an upward curvature.
This kind of effective exponent behavior usually corresponds to logarithmic 
correction to scaling. In case of the general form 
\begin{equation}
\rho(\infty) = \epsilon^{\beta} \ln^{x}(\epsilon) \ , 
\end{equation}
where $\epsilon=|p-p_c|$ the effective exponent behaves as
\begin{equation} \label{logform}
\beta_{eff} = \frac {d \ln( \rho(\infty) ) } {d \ln(\epsilon) } = 
\beta + \frac {x} { \ln (\epsilon) }  \ . 
\end{equation}
Applying this form for the data of Fig.\ref{beta3432_18} one obtains:
$\beta=1.05(5)$, $x=1.2(1)$ for $D=0.1$ 
and $\beta=0.93(10)$, $x=-1.1(1)$ for $D=0.8$. Although the assumed 
scaling correction form may look somewhat ad hoc it indicates $d_c=1$.

I also considered the scenario suggested by \cite{DCM05} according to
which PCPD scaling ($\alpha=0.19(1)$ \cite{pcpd2cikk,DCM05}) should be 
observed at the transition point. By assuming that such density decay 
appears for very long times ($t > 10^9$ MCS) one can read-off the
corresponding slightly different critical point estimates:
$p_c=0.30325(5)$ for $d=0.1$ and $p_c=0.4071(1)$ for $D=0.8$.
Using these values in the local slopes analysis one can obtain: 
$\beta=0.85(5)$ for $D=0.1$ and $\beta=0.80(5)$, neither of them is 
near to the exponent $\beta=0.40(1)$ of the PCPD class \cite{pcpd2cikk}.
Therefore these simulations can't support the PCPD class scenario.

\section{Simulations of the $3A\to 4A$, $3A\to\emptyset$ model} 
\label{3430sec}

Considering the mean-field type of scaling behavior of the 
$3A\to 4A$, $3A\to 2A$ model one may speculate that in this model the
spatial fluctuations are somewhat suppressed: it's hard to create nearly
arbitrarily large regions void of particles from a place where
"annihilation" reactions have taken place due to the $3A\to 2A$ rule. 
In the low-diffusion regime, when a particle was created from the 
configuration $...0AAA0...$ it is most likely that this offspring would 
recombine with the remaining $2A$ particles and undergo another local 
sequence of reactions. And if in the meantime the $3A$ in a row have 
either branched or undergone a $3A\to 2A$ reaction, the remaining
particles would not have the time to go very far. Since everything
takes place essentially locally, the mean-field rate equation should be
valid \cite{Dorpriv} yielding therefore $\alpha=1/3$.

To check this scenario I run simulations for the $3A\to 4A$, $3A\to\emptyset$ 
model, in which no such local sequence of reactions occur, since following 
the triplet annihilation the remaining single $A$ can diffuse away at most.

The dynamical rules are very similar to those of the 
$3A\to 4A$, $3A\to\emptyset$ model.
An elementary MCS consists of the following processes:
\begin{description} 
\item[(a)] $A\emptyset\leftrightarrow\emptyset A$ with probability D,
\item[(b)] $3 A \to\emptyset$ with probability $p(1-D)$,
\item[(c)] $3 A \to 4A$ with probability $(1-p)(1-D)$,
\end{description}
such that the reactions were allowed on the left or right side of the 
selected particle strings. Now the system size was $4\times L=10^5$ 
with periodic boundary conditions and the density of particles, singlets
and triplets is followed up to $2\times 10^9$ MCS from random initial state.
\begin{figure}
\epsfxsize=70mm
\epsffile{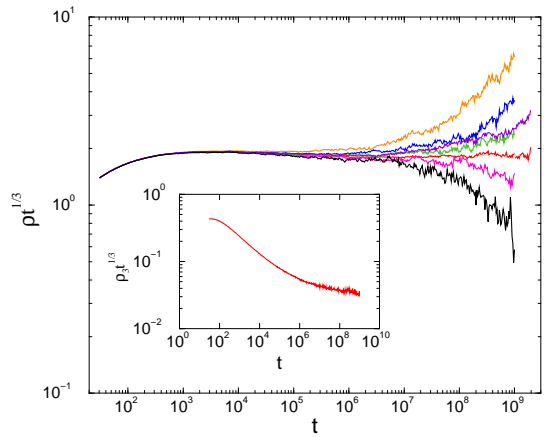}
\vspace{2mm}
\caption{(Color online)
Density decay in the $3A\to 4A$, $3A\to\emptyset$ model at $D=0.8$.
Different curves correspond to $p=0.1185$, $0.1187$, $0.11875,
$$0.1188$, $0.11886$, $0.1189$, $0.119$ (from top to bottom). 
The insert shows the decay of triplets for $p=0.11886$.}
\label{3430_8}
\end{figure}
As Figure \ref{3430_8} shows the time evolution of $\rho(t)$ can
be split into three parts (a) an initial slow regime ($t<3\times 10^3$ MCS),
(b) an intermadiate faster then mean-field regime 
($3\times 10^3 < t < 3\times 10^6$ MCS), (c) a mean-field regime for
$t>3\times 10^6$ MCS. One can see a level-off in the time evolution of
$\rho t^{1/3}$ for $p=p_c=0.11887(2)$. 
For this critical $p$ value the density of triplets 
behaves in the same way in the long time limit ($t >\simeq 3\times
10^6$ MCS) corroboting
the $d_c=1$ result of the previous section. 

Finally the static exponent $\beta$ was determined in the active phase
in the neighbourhood of the critical point ($p_c=0.11887$). 
As Figure \ref{beta3432_18} shows the local slopes (\ref{beff})
converge to the mean-field value again. By fitting with the form
(\ref{logform}) one gets the following estimates: 
$\beta = 0.99(5)$, $x=0.6(1)$.

\section{Conclusions}

Large scale simulations for two different triplet models: 
$3A\to 4A$, $3A\to 2A$ and $3A\to 4A$, $3A\to\emptyset$ result in
mean-field type of static and dynamic scaling behavior in one dimension.
The simulations up to $t=2\times 10^9$ MCS do not support the scenario 
according to which fluctuations are supressed and reactions take place locally 
hindering to see PCPD type of critical behavior. Since the triplet density 
dacays in the same way as the total density, which is typical at the 
upper critical dimension the $d_c=1$ is concluded. Furthermore logarithmic
correction to scaling is shown in case of the static order parameter
exponent. On the other hand one can't see log. corrections in the
density decay, which may mean that these corrections are small 
or perhaps a next to the leading order correction term hinders to see 
it. Such correction term was found in a very recent field theoretical 
analysis of the $3A \to (\emptyset,A,2A)$ models \cite{LG06}, which 
correspond to the dominant behavior in the inactive phase of the 
triplet model I studied. 

The contradiction with the results of the Langevin equation analysis
of a bosonic triplet system is unresolved. This may also mean that the 
two-species coupled model and the one-species model behave differently.

\vskip 0.5cm

\noindent
{\bf Acknowledgements:}\\
I thank I. Dornic and M. Mu\~noz for the useful discussions.
Support from Hungarian research funds OTKA (Grant No. T046129) is acknowledged.
The author thanks the access to the NIIFI Cluster-GRID, LCG-GRID 
and to the Supercomputer Center of Hungary.

\end{document}